# Computationally efficient human body modelling for real time motion comfort assessment

Raj Desai, Marko Cvetković, Junda Wu, Georgios Papaioannou, Riender Happee

Cognitive Robotics (CoR) - TU Delft
r.r.desai@tudelft.nl

**Abstract.** Due to the complexity of the human body and its neuromuscular stabilization, it has been challenging to efficiently and accurately predict human motion and capture posture while being driven. Existing simple models of the seated human body are mostly two-dimensional and developed in the mid-sagittal plane exposed to in-plane excitation. Such models capture fore-aft and vertical motion but not the more complex 3D motions due to lateral loading. Advanced 3D full body active human models (AHMs), such as in MADYMO, can be used for comfort analysis and to investigate how vibrations influence the human body while being driven. However, such AHMs are very time-consuming due to their complexity. To effectively analyze motion comfort, a computationally efficient and accurate three dimensional (3D) human model, which runs faster than real time, is presented. The model's postural stabilization parameters are tuned using available 3D vibration data for head, trunk and pelvis translation and rotation. A comparison between AHM and EHM is conducted regarding human body kinematics. According to the results, the EHM model configuration with two neck joints, two torso bending joints, and a spinal compression joint accurately predicts body kinematics.

**Keywords:** MADYMO, multibody model, motion comfort, posture, vibrations.

## 1 Introduction

Automated driving (AD) holds great promise to provide safe and sustainable transport. Automation will allow users to take their eyes off the road, freeing up time for work or leisure activities. However, such conditions will further complicate occupants' overall postural stability compared to conventional vehicles [1]. Being exposed to whole-body vibration (WBV) produced by the vehicle, occupants may feel discomfort while such vibrations could even cause low back pain and injuries in the lumbar spine [2]. Hence, knowledge and models of human motion and perception are essentially needed for human centred design of automated driving systems [3]. Therefore, it is important to investigate how vibrations are transmitted through the human body and how the human body responds to WBV. The most effective technique to comprehend human motion is to collect data from experiments, however this method sees several drawbacks. While tests can offer insightful information about human motion, they might not be able to record the whole spectrum of motion [4] as models can do. Some of the most advanced human body models are THUMS [5] and Simcenter Madymo AHM [6]. However, these require large amounts of computational time. Thus, the need to develop a computationally efficient and accurate 3D human body model has risen with a potential application in a variety of domains.

The proposed computationally efficient model is based on MADYMO's rigid body modelling features. The inertial properties of the bodies are incorporated in the rigid bodies of the model, while their geometry is described by means of ellipsoids, and planes. Structural deformation of flexible components is lumped in kinematic joints in combination with dynamic restraint models. Deformation of soft tissues (like flesh and skin) is represented by force-based contact characteristics defined for the ellipsoids. These characteristics are used to describe contact interactions within the models and between the model and the seat. In MADYMO, the MADYMO detailed active human model (AHM) represents the 50th percentile male population and has been validated for impact conditions [7], [8] and for vibration and dynamic driving [9]. The model geometry consists of standing height (1.76m), sitting height (0.92m) and weight 75.3 kg derived from the ergonomic model in RAMSIS [3], [10]. The AHM includes controllers to stabilise the spine, neck, shoulders, elbows, hips, knees, while it consists of 190 bodies (182 rigid bodies and 8 flexible bodies) and finite element (FE) surfaces capture the skin for contact interaction. Due to the above, the AHM requires significant amounts of computational time. To reduce this for vehicle comfort simulation, we present a computationally efficient human model (EHM) for comfort analysis.

The EHM model is designed to be computationally efficient and simple, while accurately representing body joint biomechanics and providing a good fit with experimental motion [4]. A functional set of body segments is used in the model's construction, including only those which have major influence on body kinematics and dynamics. The model shall represent seat interactions as a dependent function of posture. Hence, the model has realistic contacts with floor, seat base and seat back. To benchmark the EHM, the AHM is used for comparison of model performance. We validate both models for head, trunk and pelvis motion in translation (x-y-z) and rotation (roll-pitch-yaw) in response to 3D seat motion (x-y-z). To our knowledge there is no thoroughly (6 DoF with head, pelvis and trunk, vertical/fore aft/lateral) validated body 3D multibody human body model reported in literature. To address these challenges, this paper develops an efficient multibody (MB) seated human body model that can be used for predicting human body response in a dynamic driving scenario.

## 2 Biomechanical modelling

In order to build an efficient seated human body model, models in literature [11], [12] and the MADYMO AHM were investigated. As described above we adopted a body size matching the AHM but reduced the model complexity[4]. Fig. 1 shows the EHM and the AHM in the configuration used for validation with experimental data [4] . This was performed with an experimental seat with two configurable back supports. From this dataset we selected the condition with high support and erect posture. The lower support pad was at the posterior superior iliac spine, and the higher support pad was aligned with the apex of the scapula's angulus inferior.

The EHM consists of 12 segments: pelvis, lower torso, middle torso, upper torso, neck, head, left thigh, right thigh, left lower leg, right lower leg, left foot and right foot. The inertia of the arms is incorporated in the upper torso segment. The different body segments are connected by kinematic joints. The lumbar and thoracic spine includes three joints, where the lowest joint is locked since the current data could well be captured with two joints. The lowest joint (now locked) connects pelvis and lower torso. An additional spherical joint is placed between L4-L5 to capture lumbar bending [13] as this forms the rotational point between lower and middle torso. The middle and upper torso are connected by a spherical-translational joint, which allows 3D rotational and vertical motion. This vertical motion is essential in capturing spinal compression/extension in vertical loading. The spherical joints are used to model 3D rotation capturing flexion-extension, abduction-adduction and yaw rotation of the torso. The cervical spine includes two joints. A spherical joint is placed at the upper neck located at (C1-C0) to capture the head yaw-pitch-roll and at the lower neck (T1-C7) universal joint to capture the roll-pitch motion [14]. The right and left hip joints are also modeled as spherical joints to connect the thighs and pelvis, while the right and left knee joints are modeled as revolute joints, allowing for relative rotational movement around one axis between the thighs and lower legs. Ankles connections are modelled as revolute joints, and they connect lower legs and feet. The center of gravity (CoG) for the head, trunk (the eighth thoracic vertebra - T8), and pelvis in the EHM is located similarly to that of the AHM. More details about the joints are shown in Table 1. The EHM model has 31 degrees of freedom (DoF) considering the various rotational and translational movements allowed by the kinematic joints and their constraints.

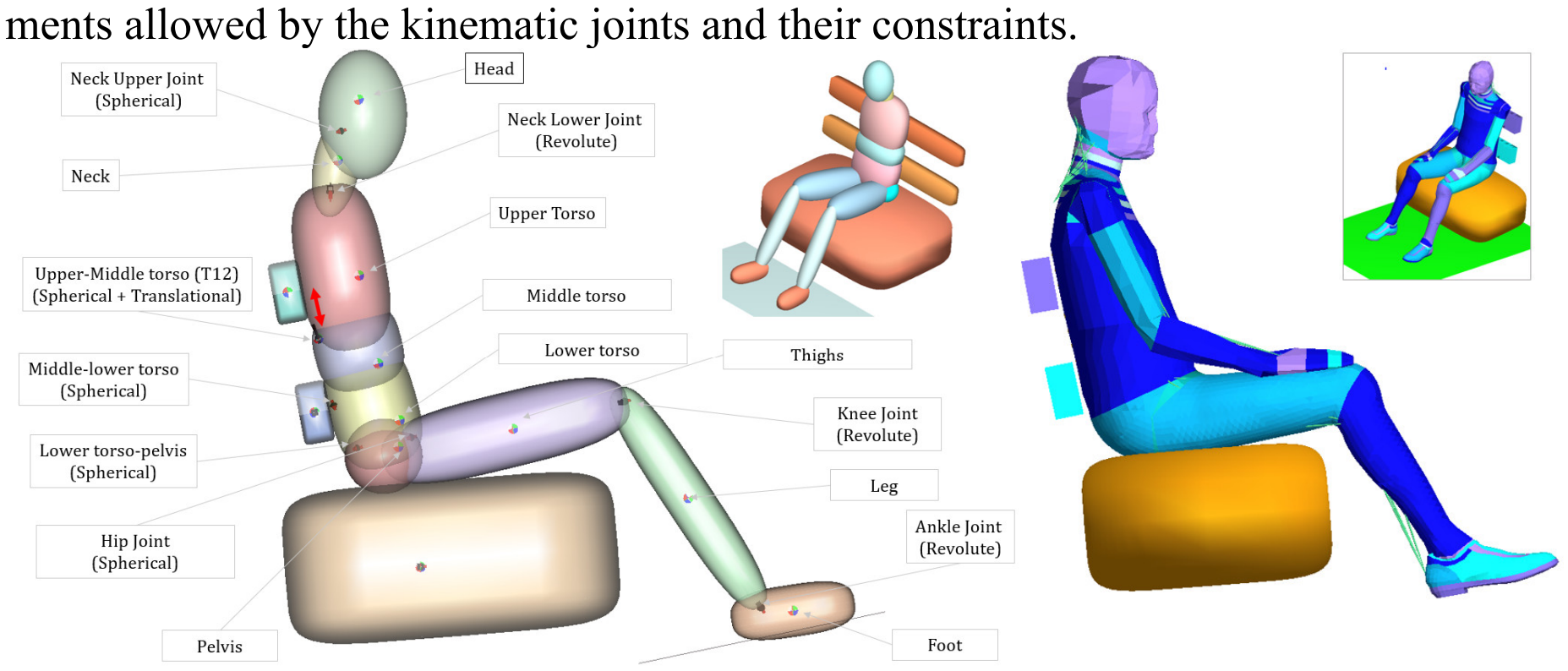

Fig. 1. Human body modelling. (Left: EHM, Right: AHM, Top corner: Isometric view)

To capture the human-seat interaction, MB-MB contacts have been established. Thus, human body contacts with seat cushion, seatback and floor are defined. Contact interactions are specified using a master surface against a slave surface. Selected groups of multibody surfaces are used as master (planes and ellipsoids) and slave (ellipsoids) surface, such as feet contact with floor, pelvis with seat pan and torso with backrest. In this model, all possible contacts and the contacting surfaces can penetrate each other. The corresponding elastic contact force depends on the penetration. The deformation of muscle, fat and skin are captured using point restraints with linear stiffness and damping and contact surfaces. These are currently defined as linear force-deflection characteristics with stiffness in N/m and damping in Ns/m. To realistically predict how a seated human body responds to vehicle vibration, feet, legs and their contact with the floor are added, since our previous research has shown relevant contributions of the legs in trunk stabilization in a dynamic slalom drive [9].

The deformation of bony segments and multiple body joints are lumped into joint compliance models. In complex biomechanical models, the intervertebral joints, are modelled as 6 DoF joints allowing compression, shear, and rotation, requiring the tuning of many parameters. In the EHM, joints are efficiently modelled, by reducing their DoFs, to achieve increased prediction capabilities and keep computational cost to the minimum. In this direction, most bodies are interconnected by spherical or revolute joints rather than linear (translational) springs and dampers. Complex human models include passive joint resistance models complemented with models of active muscular joint stabilization [15]. Such models include both reflexive (feedback) stabilization and joint "stiffening" through co-contraction of antagonist muscle groups. To simplify the model and the estimation of the parameters we defined linear stiffness and damping for each joint degree of freedom. This linear stiffness and damping was initially defined using the restraint cardan feature. The Cardan restraint consists of three torsional parallel springs and dampers that connect two bodies. The torques depend on the Cardan angles that describe the relative orientation of the corresponding restraint coordinate systems.

Tuning the model, low stiffness values were found in particular in the neck. This led to substantial postural drift, with the actual posture deviating from the specified posture. This was resolved replacing cardan restraints by PID controllers where integral action eliminated drift but hardly affected the dynamic response. The proportional action replaced the cardan stiffness, and the differential action replaced the cardan damping [9].

Table 1. Model joints configuration. (T: Torso, Trans: Translation in vertical direction)

| Joint | Connected body segments | | Joint type | Co-ordinates (cm) | | |
|---|---|---|---|---|---|---|
| | | | | *x* | *y* | *z* |
| Upper Neck($J_1$) | Head | Neck | Spherical | -8.85 | 0 | 63.04 |
| Lower Neck ($J_2$) | Neck | Upper T | Universal | -10.85 | 0 | 51.39 |
| T12 joint ($J_3$) | Upper T | Middle T | Spherical + Trans | -16.01 | 0 | 22.90 |
| L4-L5 joint ($J_4$) | Middle T | Lower T | Spherical | -12.50 | 0 | 8.14 |
| Waist joint ($J_5$) | Lower T | Pelvis | Spherical (Lock) | -7.95 | 0 | -0.21 |
| Hip joint right ($J_6$) | Pelvis | Thigh right | Spherical | 2.54 | 8.86 | 1.78 |
| Hip joint left($J_7$) | Pelvis | Thigh left | Spherical | 2.54 | -8.86 | 1.78 |
| Knee right ($J_8$) | Thigh right | Leg right | Revolute | 43.08 | 14.01 | 8.65 |
| Knee left ($J_9$) | Thigh left | Leg left | Revolute | 43.08 | -14.01 | 8.65 |
| Ankle Right ($J_{10}$) | Leg right | Foot right | Revolute | 69.40 | 15.75 | -30.67 |
| Ankle left ($J_{11}$) | Lower leg | Foot left | Revolute | 69.40 | -15.75 | -30.67 |

Moments of inertia are calculated according to the models shape of each segment (Table 2). In the EHM model, the positioning of the joints and the location of the CG for each segment are critical considerations for the realistic prediction of the dynamics and kinematics in the model. Proper joint positioning ensures that the model can replicate the expected range of motion and joint behavior, while accurate CG location for each segment helps to capture the segmental mass distribution and its effect on overall motion dynamics. These factors are carefully considered in the development of the EHM model. The efficient model is overlaid with the AHM to encapsulate the symmetric nature and true anthropometric characteristics of the human body.

Table 2. Model body segment data.

| Body | Ellipsoid Degree | Mass | Moment of Inertia (kg·m$^2$) | | | Co-ordinates (cm) | | |
|---|---|---|---|---|---|---|---|---|
| | | | $I_{xx}$ | $I_{yy}$ | $I_{zz}$ | $x$ | $y$ | $z$ |
| Head | 2 | 6.23 | 0.031 | 0.031 | 0.020 | -5.461 | 0 | 69.104 |
| Neck | 2 | 1.6 | 0.003 | 0.004 | 0.005 | -9.420 | 0 | 57.393 |
| Upper torso | 3 | 8.93 | 0.238 | 0.146 | 0.181 | -8.231 | 0 | 34.692 |
| Middle torso | 3 | 7.7 | 0.238 | 0.146 | 0.181 | -3.827 | 0 | 15.927 |
| Lower torso | 3 | 10.70 | 0.137 | 0.078 | 0.117 | 0.2164 | 0 | 5.172 |
| Pelvis | 2 | 10.93 | 0.115 | 0.050 | 0.151 | 0 | 0 | 0 |
| Thighs right | 3 | 7.7 | 0.007 | 0.129 | 0.129 | 21.859 | 11.24 | 3.450 |
| Thighs left | 3 | 7.7 | 0.007 | 0.129 | 0.129 | 21.859 | -11.24 | 3.450 |
| Legs right | 3 | 3.58 | 0.031 | 0.031 | 0.020 | 55.405 | 14.8 | -9.823 |
| Legs left | 3 | 3.58 | 0.031 | 0.031 | 0.020 | 55.405 | -14.8 | -9.823 |
| Feet right | 3 | 1.116 | 0.001 | 0.005 | 0.004 | 75.409 | 16.59 | -31.332 |
| Feet left | 3 | 1.116 | 0.001 | 0.005 | 0.004 | 75.409 | -16.59 | -31.332 |

## 2.1 Parameter identification

A simple seat experiment conducted by our research group is used to validate the model [4] and improve its fitting by optimizing selected parameters. In the experiment, participants sat in a car mock-up and were excited with random vibrations (0.3 m/s$^2$ rms) in vertical, fore-aft and vertical directions. Similarly with the car mock-up, the MADYMO model environment consists of three segments: seat pan, backrest and the floor. The floor is a plane and other segments are ellipsoids. A 50th percentile male body size is adopted and its mass is close to the average human model to facilitate comparison. These data were obtained from anthropometry measurements in literature [16]. Thus, some of the human body parameters (i.e., mass and inertia values) are predefined, while others (i.e., translational-rotational spring and damping) are to be determined. More specifically, each spherical joint has 3 DoF, each revolute or translational joint has 1 DoF. Each DoF is modelled by one linear spring and one linear damper. Therefore, every one DoF corresponds to two design parameters. The stiffness and damping represent passive tissue resistance, as well as postural stabilization using muscle feedback and co-contraction. The AHM is enabled with posture controllers to stabilize the body, while a simpler approach is adopted in EHM to reduce computation time and the number of parameters. This study will illustrate how such a simplification affects accuracy.

To validate the models, the predicted responses are compared to experimental data. There can be significant individual variation in muscle activation patterns and joint stiffness, which make challenging the development of models that accurately predict the individual body kinematics. Currently, the model is configured to represent the average response over the group of participants. The model should be able to capture the experimental response accurately for head, pelvis, trunk in vertical, fore-aft and lateral directions. As response functions, specific gains of different body segments will be evaluated using transfer functions in the frequency domain, as defined below:

$$\text{Gain} = \frac{\mathcal{F}(s_o)}{\mathcal{F}(s_i)} \tag{1}$$

wherein $s_o$ is human response of specific body segment in time domain, such as pelvis's vertical displacement or head's pitch; $s_i$ stands for input vibration in time domain; $\mathcal{F}$ stands for Fourier transform, which means the gain is a function in frequency domain The relevant gains should have minimum errors with respect to experimental data in different seat motions, for the EHM to accurately predict the human response. Therefore, these errors of specific gains are the criterion, i.e., the cost function, for parameter identification. The Butterworth band pass filter of 0-12 Hz frequency range is applied in both experimental as well as model responses to isolate signals within this range. In each seat motion, the gain of experiment as function of frequency is denoted as $Gain_{exp}$ and the gain of model as function of frequency is denoted as $Gain_{model}$. In order to bring the gain value on the same scale for all different gains, relative gain expression is formulated. The relative gain is denoted as $Rel_{Gain}$ and is formulated as:

$$Rel_{Gain} = Gain_{exp} + 0.05\overline{Gain}_{exp} \quad (2)$$

Due to the application of this model (i.e., comfort analysis), the priority is the accurate prediction of human response in low frequencies (0.1-3Hz). Thus, we prioritize this by using weighting factors. This is captured by dividing with corresponding square of frequencies. The error criterion for each individual objective function is formulated as:

$$Crit_1 = RMS\left\{\frac{Gain_{exp} - Gain_{model}}{Rel_{Gain} f^2}\right\} \quad (3)$$

The required body segments for model fitting are head, upper torso and pelvis.

## 3 Numerical Simulations

To replicate the experimental conditions the same input signal was used as experienced by the participants. The seat received vibrational input in the x, y, and z directions. The vibrational excitations to the seat are given thereafter. The co-simulation and optimization flow chart between MATLAB-SIMULINK-MADYMO is shown in Fig. 2. The same process was applied to the AHM. The EHM used ellipsoid-ellipsoid contacts for the seat back whereas the AHM used a solid FE model for the seat back foam. At the point of contact, shear forces are facilitated by the FE backrest. The integration time step size was set to 1E-3 s for EHM whereas due to presence of FE and detailed multibody components a smaller time step of 5E-5 s was adopted in the AHM.

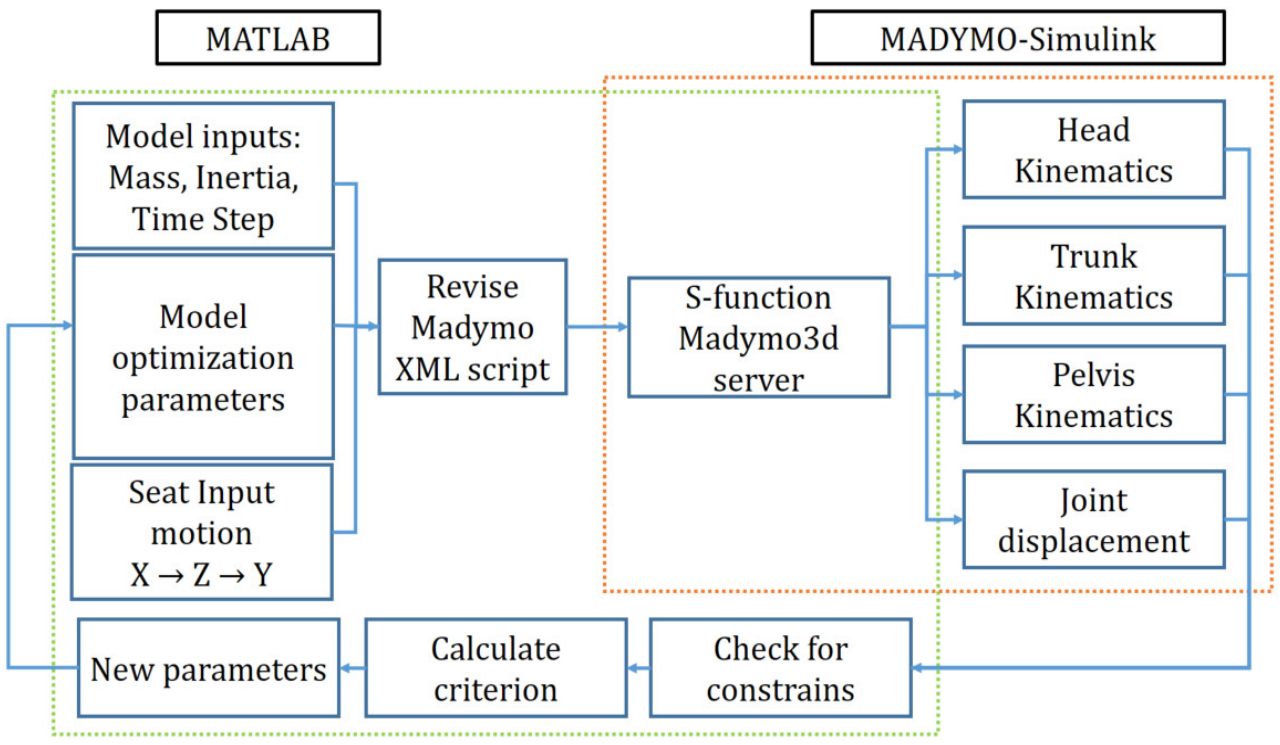

Fig. 2. Co-Simulation flow chart

# 4 Results

The experimental and model results are presented in Fig. 3-5. During the optimization of the model parameters, the accuracy in capturing the head and trunk motion was prioritized over pelvic movements due to complexity of the pelvis and its interactions with other parts of the body. With the optimization, the AHM and EHM accurately capture the experimental response, while both models demonstrate a higher accuracy in capturing the gain of the head and trunk movements compared to the pelvis movements in the experimental data. The EHM outperforms the AHM in certain sets of experimental data.

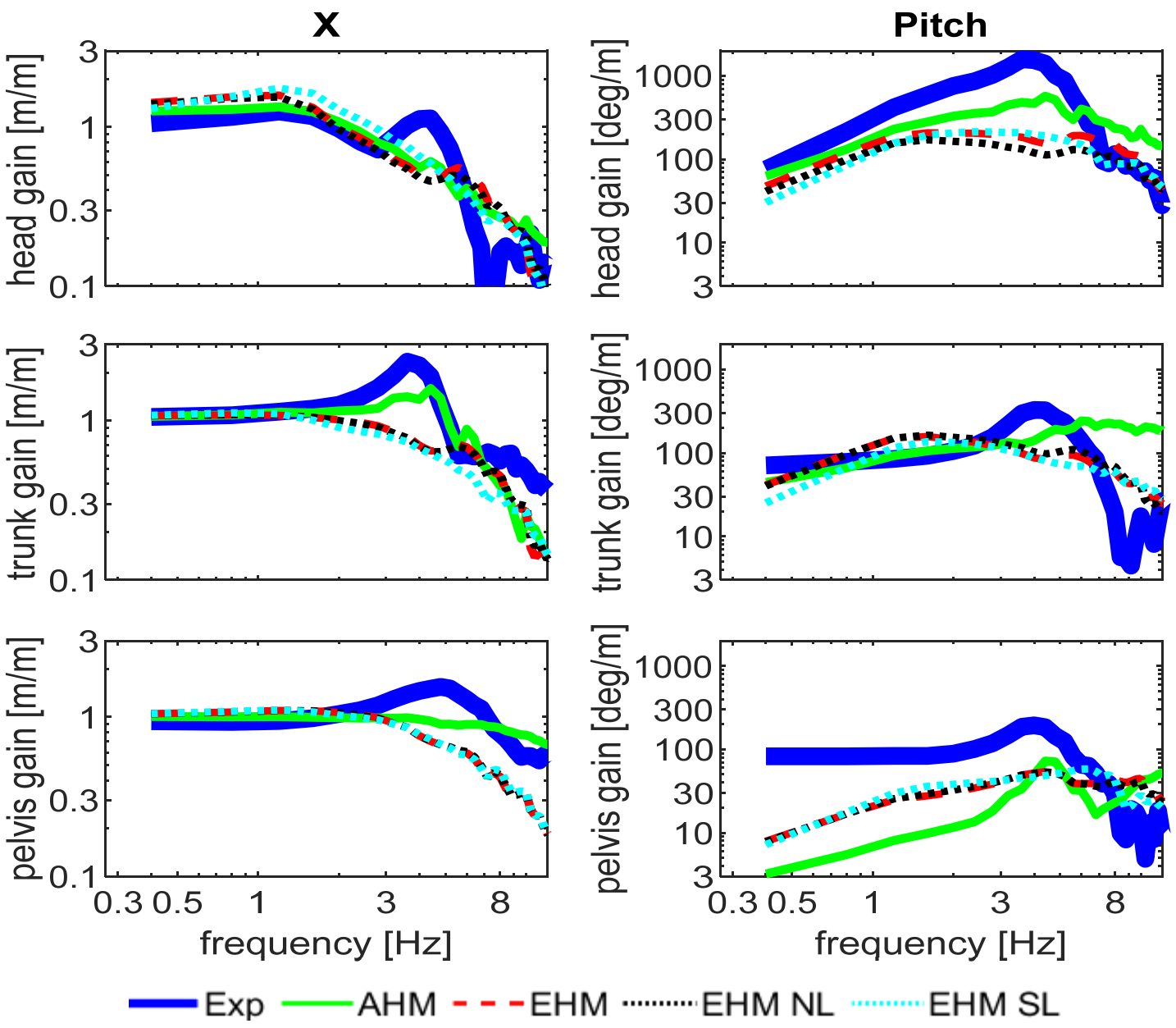

Fig. 3. Model simulation results fore-aft loading case.

In this paper, the EHM is also simulated locking selected joints in order to investigate the requirements of two neck and spinal bending joints. By specifically locking the T12 location spherical and compression joint spine locked (SL), there is a substantial impact on pelvis, trunk and head in vertical loading, where the oscillation frequency increases to an unrealistic peak at 8 Hz with locked spine. This shows the importance of the vertical compression joint in the lumbar spine. Locking the lower neck (NL) slightly reduced head pitch in fore-aft and vertical loading. However locking the lower neck joint had a much more profound effect in lateral loading, where the model with one neck joint highly underestimated head roll and yaw. This shows that a two-joint neck model is particularly relevant for lateral loading. This emphasizes the importance of considering at least two neck and spine joints, in biomechanical simula-tions and movement control studies. This model also offers a comprehensive approach for capturing intricate interactions between the human body and a seat, with potential applications in ergonomic seating design and musculoskeletal disorder prevention. Future work includes implementing active muscle controllers to capture reflexes, postural adjustments, and advanced feedback models such as proprioceptive, vestibular, and visual motion perception. This will enable the design of innovative control algorithms for automated vehicles, ultimately improving comfort during automated driving.

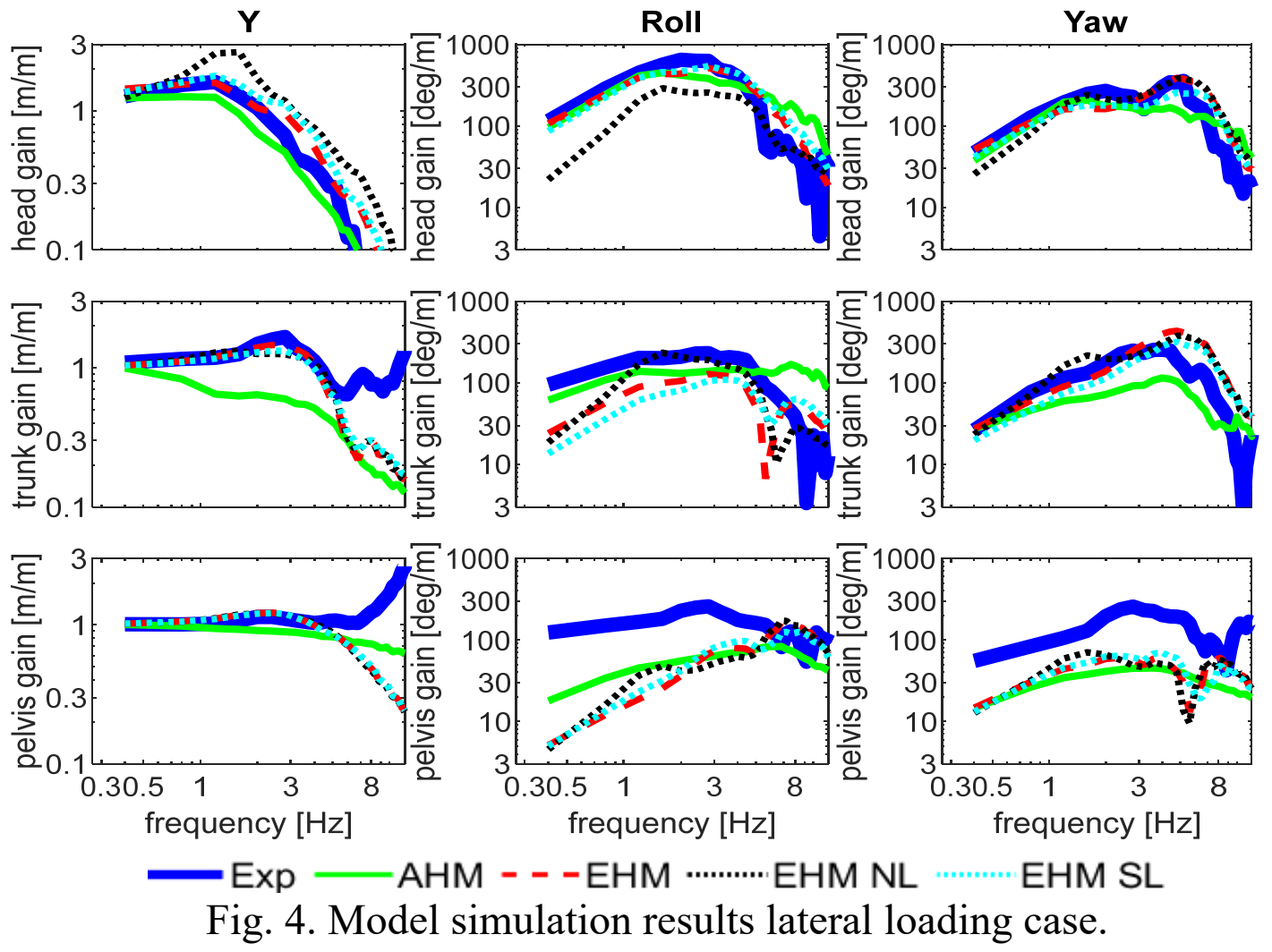

Fig. 4. Model simulation results lateral loading case.

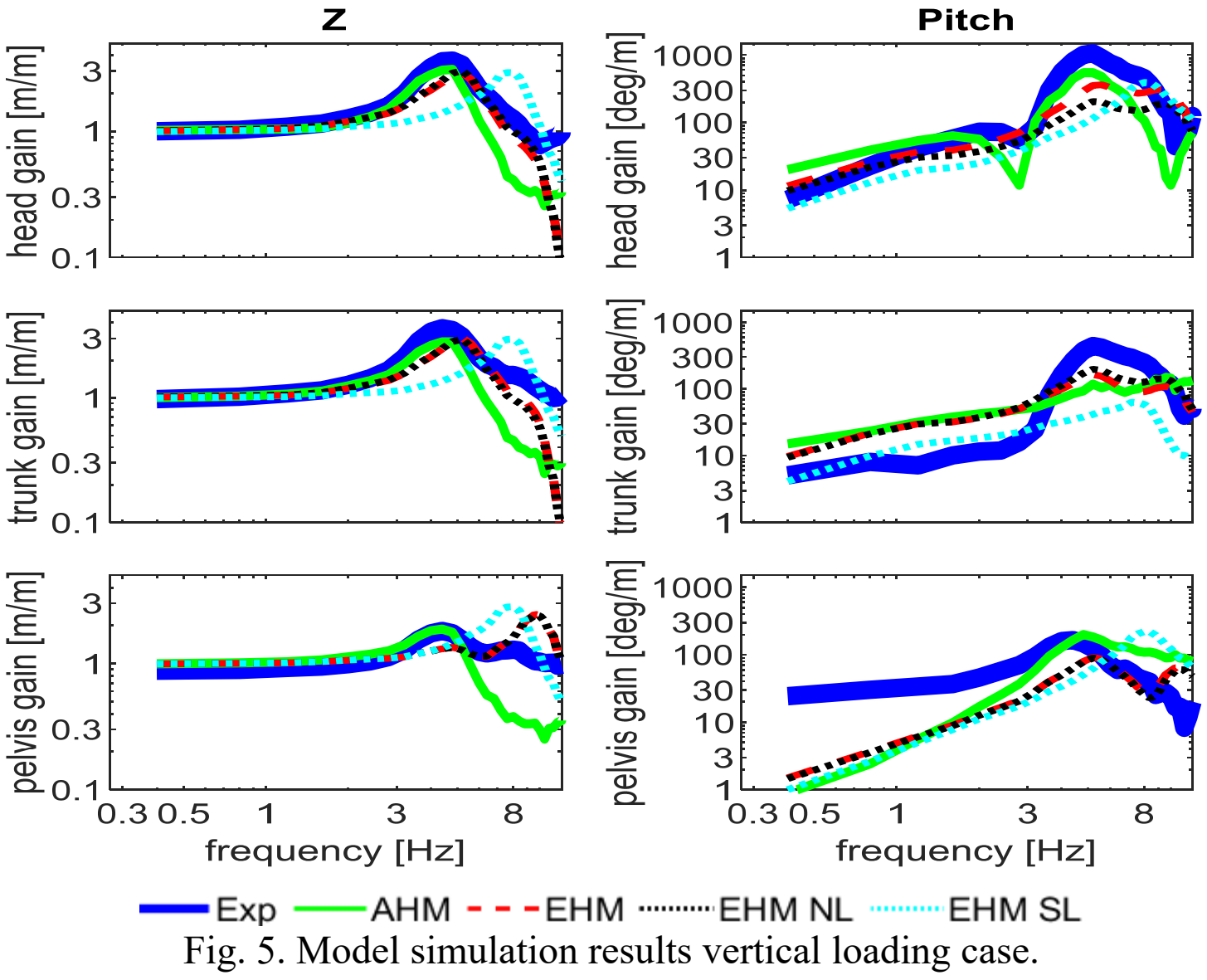

Fig. 5. Model simulation results vertical loading case.

# 5 Conclusion

A validated multibody human body model with realistic joint configurations, capturing compression/shear contact interactions using MB-MB contact with friction, is developed. The model is highly efficient and accurate, demonstrated through comparison with the complex MADYMO active model using rich experimental data. The paper also describes a reliable procedure to estimate the human model and contact parameters to fit experimental human response data. The fitting methodology and computational efficiency of the model pave the way to develop individual postural stabilization models. These individual models can be adapted to the individual anthropometry [17] and postural control strategies. Compared to AHM, the EHM gets similar or even better human responses in some cases with regards to the experimental data. The EHM is much faster (30 sec) than the AHM (3 hr) for completion of 35 sec long of simulations. The arrangement of two neck joints, two torso bending joints, and a spinal compression joint was found to be sufficient in accurately capturing 3D body kinematics.

**Acknowledgement**: We acknowledge the support of Toyota Motor Corporation.